# Translatory and rotatory motion of Exchange-Bias capped Janus particles controlled by dynamic magnetic field landscapes

*Rico Huhnstock\*[1),2)], Meike Reginka[1),2)], Andreea Tomita[1)], Maximilian Merkel[1),2)], Kristina Dingel[2),3)], Dennis Holzinger[1)], Bernhard Sick[2),3)], Michael Vogel[1),2)], Arno Ehresmann[1),2)]*

1) Institute of Physics and Centre for Interdisciplinary Nanostructure Science and Technology (CINSaT), University of Kassel, Heinrich-Plett-Strasse 40, D-34132 Kassel, Germany
E-Mail: rico.huhnstock@physik.uni-kassel.de, Phone: +49 561 804-4015

2) Artificial Intelligence Methods for Experiment Design (AIM-ED), Joint Lab of Helmholtzzentrum für Materialien und Energie, Berlin (HZB) and Kassel University, Germany

3) Intelligent Embedded Systems, University of Kassel, Wilhelmshöher Allee 71-73, D-34121 Kassel, Germany

ABSTRACT. Magnetic Janus particles (MJPs), fabricated by covering a non-magnetic spherical particle with a hemispherical magnetic in-plane exchange-bias layer system cap, display an onion magnetization state for comparably large diameters of a few microns. In this work, the motion characteristics of these MJPs will be investigated when they are steered by a magnetic field landscape over prototypical parallel-stripe domains, dynamically varied by superposed external magnetic field pulse sequences, in an aqueous medium. We demonstrate, that due to the engineered magnetization state in the hemispherical cap, a comparably fast, directed particle transport and particle rotation can be induced. Additionally, by modifying the frequency of the applied pulse sequence and the strengths of the individual field components, we observe a possible separation between a combined or an individual occurrence of these two types of motion. Our findings bear importance for lab-on-a-chip systems, where particle immobilization on a surface via analyte bridges shall be used for low concentration analyte detection and a particle rotation over a defined position of a substrate may dramatically increase the immobilization (and therefore analyte detection) probability.

Magnetic micro- and nanoparticles are considered to be important tools for the realization of various life and bio science applications and technologies[1,2]. In lab-on-a-chip systems (LOC) or micro-total-analysis-systems (µTAS) magnetic particles may constitute a key component since a controlled particle actuation allows for the implementation of a variety of functionalities like fluid mixing, sorting, drug uptake and delivery as well as detection of analyte biomolecules[3,4]. One of the analyte detection strategies is the dimerization of particles or the immobilization of particles on a substrate via analyte bridges. These bridges are formed by binding of an analyte molecule to two molecular counterparts, one residing on the particle surface, the other residing on another particle or on the chip surface. It is evident that the binding sites must be in close proximity to each other for binding events to happen. Therefore, a controlled approach of the binding sites is necessary. As the area occupied by a molecular binding site is typically much smaller than the surface of micron sized particles, not only a translational particle motion is necessary, but also a controlled particle rotation. In particular when analyte concentrations are low and the particle surface is not fully covered by bound analytes, a particle rotation over a specific chip surface position will enhance the probability for a binding event and particle immobilization. For superparamagnetic particles, which are discussed for the use in LOC or µTAS systems in the vast majority of cases, a controlled physical rotation by a rotating magnetic field is not possible. For a rotatory motion, magnetic Janus particles (MJPs) may be used instead. These particles consist of two sides with different physical characteristics[5], where, *e.g.*, one part of the particles has been functionalized magnetically for inducing particle motion and the other part with analyte sensitive reagents for analyte uptake[6,7]. The potential of this approach has been demonstrated by the activation of T cells, where magnetic MJP have been remotely steered towards them enhancing the particle-cell recognition through a controlled spatial rotation of the particles due to an applied external magnetic field[8]. Further works highlight the flexibility of externally controlling magnetic MJP rotation due to their asymmetric structure[9]. Besides other known fabrication methods for MJPs[10–12], the magnetic functionalization can be achieved by directly depositing a magnetic thin film system on top of self-assembled, non-magnetic spheres[13] leading to the formation of a magnetic cap covering one half of the original particle[14–16]. The magnetic ground state of these caps (vortex, onion or out-of-plane) can be controlled by variation of the particle size, cap thickness and material composition[14,17]. MJPs with vortex and out-of-plane magnetic states within the caps were introduced in previous works to demonstrate a controlled steering of the MJP's movement direction whilst being transported by different propulsion mechanisms in a liquid environment[18–24]. What is missing to date is an investigation of the translational and rotational dynamics of micron-sized MJP with in-plane magnetized caps especially towards their capability for a transport to and a controlled particle rotation over specific substrate surface positions in physiological liquids[25].

In this work we, therefore, demonstrate a prototypical system where engineered MJPs with exchange-biased, in-plane magnetized caps perform a controlled rolling motion over or in proximity to a substrate surface in water. This is enabled by moving the MJP within a tailored magnetic stray field landscape superposed by dynamically varying external magnetic fields. The static stray field landscape possesses spatially confined, high magnetic field gradients, when compared with externally generated field gradients, thus generating a high magnetic force onto magnetic particles[26–28]. The field landscape emerges from magnetic domain walls of artificially generated magnetic stripe domains with remanent in-plane magnetizations. Such periodic domain patterns can be written deliberately into exchange-biased (EB) thin film systems by employing ion bombardment induced magnetic patterning (IBMP)[29–33]. The macroscopic magnetizations of the domains utilized in this work point alternatingly towards or away from each other (see Figure 1), resulting in a so-called head-to-head (hh) and tail-to-tail (tt) domain configuration. In general however, arbitrary domain patterns (periodic and non-periodic) can be fabricated using IBMP. It was shown that the combination of the resulting artificial magnetic stray fields and external magnetic field pulses alters the potential energy landscape of a superparamagnetic particle in the vicinity of the EB substrate such that a controlled and directed translatory motion can be accomplished with particle velocities up to 40 μm/s[34].

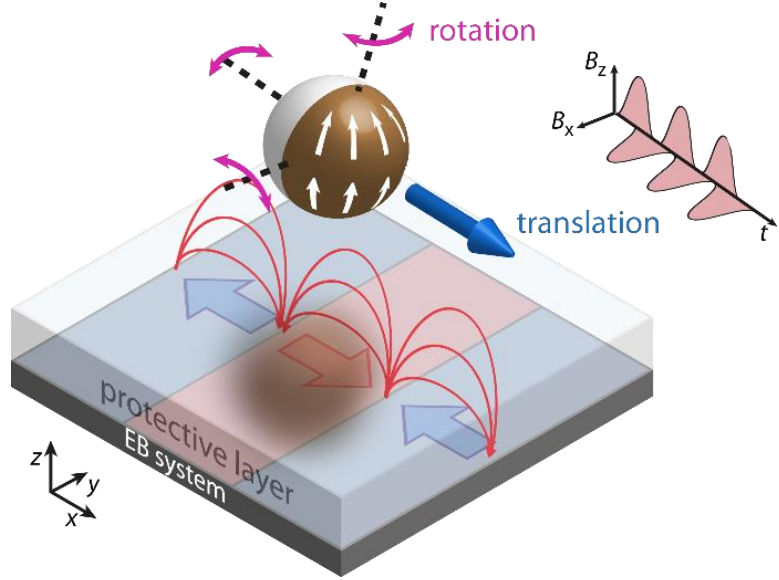

**Figure 1.** Sketch of a magnetic Janus particle (MJP), with a cap (brown) consisting of an in-plane exchange-biased layer system, interacting with an artificial magnetic stray field landscape on top of a topographically flat substrate. The stray fields are induced by an underlying periodic magnetic domain pattern. Controlled translation and rotation of the MJP is achieved by applying external magnetic field pulses in *z*- and *x*-direction.

In the present work, we make use of the same technique for studying the motion behavior of magnetically engineered MJPs, which were fabricated by sputtering an EB thin film system on top of spherical silica beads. The additional unidirectional anisotropy, originating from the EB, was introduced to stabilize a magnetic onion state inside the cap in remanence, instead of a normally preferred vortex state[14,16,17] for the studied particle size. For the onion state, magnetic moments align tangentially to the cap's surface following its curvature, while the macroscopic net moment points along the EB direction (see sketch of Figure 1). Hence, an enhanced effective magnetic moment compared to the vortex state is present, allowing a faster translatory particle motion. Additionally, the fixed spatial orientation of the magnetic net moment within the cap will induce rotational movements through magnetic torque.

We will show that the combination of the distinct magnetic properties of the MJPs and a controllably varied magnetic field landscape over a topographically flat substrate leads to an intriguing combination of comparably rapid translational and rotational movements. The transport dynamics will be analyzed by identifying position and orientation of the MJPs with respect to the underlying domain pattern and the applied external magnetic field pulse sequence. Subsequently, transport and rotational characteristics will be determined from tracking data for different experimental parameters. The acquired insights demonstrate a fast, fuel-free propulsion and motion control of the MJPs by employing weak external magnetic fields (in the range of a few mT) - underlining a high potential for low power consumption lab-on-a-chip devices. Especially the inducible rotation as an additional degree of freedom for the particle motion supports applications like a sensitive detection of analyte molecules through *e.g.* particle immobilization.

*Results & Discussion:*

Exchange biased MJPs ($d = 3$ μm) with a nominal cap-pole thickness of 55 nm[16] were dispersed in distilled water and put into a microfluidic chamber on top of a magnetically patterned, topographically flat EB thin film system, which serves as the substrate. The EB substrate possesses a magnetic head-to-head (hh)/tail-to-tail (tt) parallel-stripe domain pattern (see Figure 2) with a domain width of 5 μm (periodicity of 10 μm), artificially imprinted by ion bombardment induced magnetic patterning (IBMP)[35]. As the magnetic domain pattern in the topographically flat layer system is invisible in standard optical microscopy, the lithographically patterned resist structure needed for the IBMP has been retained on top of the EB substrate for visualization of the domain pattern. Upon adding the dispersion of particles in distilled water to the microfluidic chamber, the MJPs sediment towards the substrate surface. The orientation of the particle while sedimenting has been theoretically estimated assuming a cap of constant thickness. This cap causes a shift *s* of the particles center of mass from its geometrical center towards the cap by:

$$s = \frac{m_c}{m_c + m_s} \cdot \frac{3}{8} \frac{(r_s + t_c)^4 - r_s^4}{(r_s + t_c)^3 - r_s^3}. \qquad (1)$$

Here, $m_s$ and $r_s$ represent the mass and the radius of the silica sphere, and $m_c$ and $t_c$ the mass and thickness of the metallic cap, respectively. For the used particle dimension, a shift of $s = 0.15$ μm or 10 % of the particle's radius was calculated, concluding that the particles sediment with their magnetic cap directed towards the substrate. This is also hinted at by a dark contrast filling the whole particle in microscopic observations taken with 100x magnification to characterize the initial state prior to the induced lateral motion (see Figure 2b). When approaching the substrate, the magnetic caps of the MJPs interact magnetostatically with magnetic stray fields originating from the EB substrate, leading to positioning of the particles close to a domain wall. The particle will not reside symmetrically above a domain wall, as energy minimization leads to a position where one of the magnetic poles of the particle caps points to a position of maximum magnetic stray field flux density (in the order of 1 mT according to micromagnetic simulations).

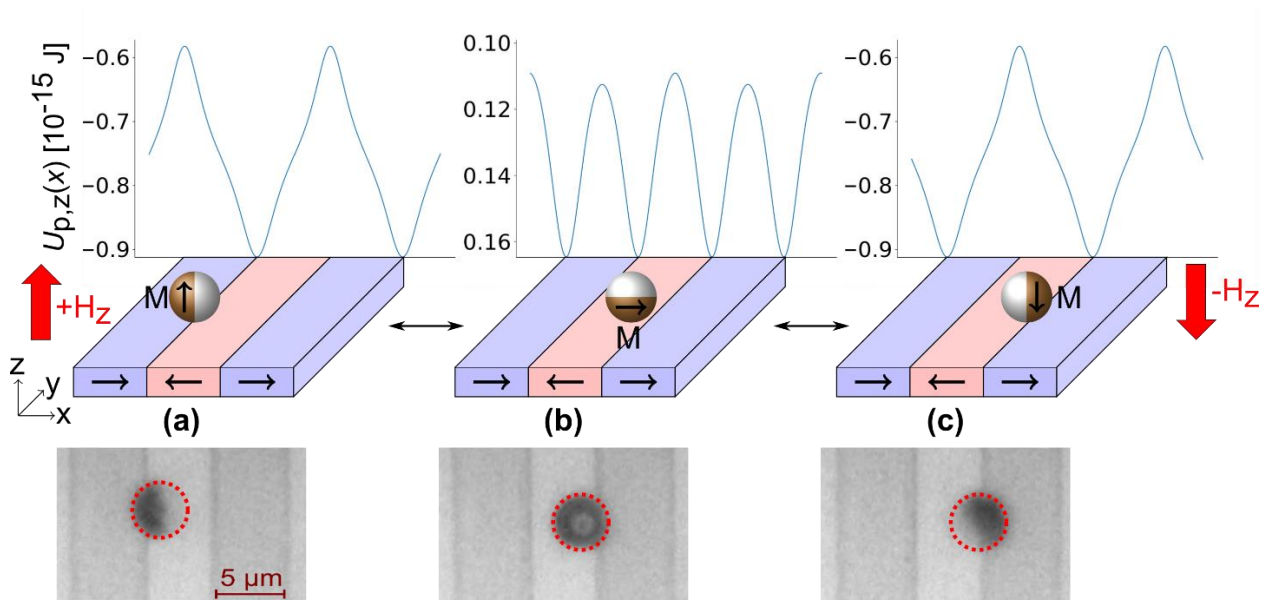

**Figure 2**. Position and spatial orientation of a magnetic Janus particle (MJP) with an in-plane exchange-bias cap in the magnetic stray field landscape of a magnetically patterned EB substrate with additionally applied field pointing into the layer plane (a), out of the layer plane (c), and with no applied external field (b). The direction of magnetization for the magnetic domains and the MJPs is indicated by arrows. The magnetic states of the MJP and its positions and orientations with respect to the underlying domain pattern with and without applied external magnetic field is sketched in the middle panels of Figs. 2 (a) to (c). Corresponding microscopy

images are shown in the lowermost panels. The calculated potential energy landscapes for the three displayed situations $U_{p,z}(x)$ are depicted as solid blue lines in the uppermost panels.

Upon applying an external magnetic field of $|H_z| \cdot \mu_0$ = 6 mT in +z- or –z-direction, a spatial displacement and reorientation of the MJP was observed. The particle as a whole moved in case of a +z-field to a location above a (hh) domain wall (Figure 2a) and in case of a -z-field to a location above a (tt) domain wall (Figure 2c). For both cases the particle cap visibly reorients in the microscopy images, seen by the changed intensity contrasts within the particles. The lateral transport of the MJP due to an applied homogenous external magnetic field $\vec{H}_{\text{ext}}$ in z-direction results from a modified magnetic stray field landscape $\vec{H}_{\text{MFL}}(x,z)$ and, thus, from an altered space dependent potential energy of the particle $U_P(x,z)$, much alike the similar transport of superparamagnetic particles[32,34]. The magnetic force $\vec{F}_M(x,z)$ exerted on the MJP can be expressed by[26]:

$$\vec{F}_M(x,z) = -\vec{\nabla} U_P(x,z) = -\mu_0 \cdot (\vec{m}_P \cdot \vec{\nabla}) \cdot \vec{H}_{\text{eff}}(x,z) \qquad (2)$$

Here $\vec{m}_P$ is the magnetic moment of the particle, i.e. the effective magnetic moment of the magnetic cap, $\mu_0$ is the vacuum permeability and $\vec{H}_{\text{eff}}(x,z)$ results from the superposition of $\vec{H}_{\text{MFL}}(x,z)$ and an externally applied, homogenous magnetic field $\vec{H}_{\text{ext}}(x,z)$. Hence, with the magnetic moment of the JP's cap aligning parallel to the effective magnetic field vector this superposition leads to the formation of a minimum in the potential energy landscape for the particle $U_P(x,z)$ above a (hh) domain wall for external +z-fields and above a (tt) domain for external –z-fields. The calculated potential energy landscape is shown by solid blue lines in the uppermost panels of Figures 2 (a) – (c). For the calculation an equilibrium distance between the particle's effective magnetic moment and the EB substrate of 1490 nm in *z*-direction (see Supporting Information S1 for further details), a particle diameter $d = 3$ µm and an effective magnetic moment $m_P = 12.46 \cdot 10^{-14}$ Am² have been used.

As observed in Figure 2a,c the resulting transport motion of the particle towards the respective locations is accompanied by a reorientation of the cap, *i.e.*, a particle rotation around the y-axis, due to the effective magnetic field exerting a torque τ according to

$$\vec{\tau} = \mu_0 \cdot \vec{m}_P \times \vec{H}_{\text{eff}}.  \quad (3)$$

The experimental data suggest that the cap orientation is mainly governed by the external magnetic field, since the torque exerted by the magnetic stray field from the substrate is not sufficient to rotate the particle. Only by applying an external magnetic field the torque is high enough to force a realignment of the magnetic cap and, hence, a rotation of the JP. Consequently, the external magnetic field serves a dual role: in superposition with the local magnetic stray fields it induces a particle translation in +x- or –x-direction and it rotates the particles. From the microscopic image, which shows the exact position of the MJP above a domain wall at approximately half the height of the particle's cap, it is suggested, that this resembles the location of the cap's magnetic center, *i.e.* the effective magnetic moment. This can be intuitively understood by considering the magnetization distribution of the attributed onion state[16]. Here, the effective magnetic moment is mainly governed by the in-plane component of the magnetization, since out-of-plane components from the two equatorial poles of the cap cancel each other. As a result of the cap's curvature the effective moment resides at a position between pole and equator.

Subsequently transport experiments for the MJPs were conducted by applying external magnetic field pulses in *x*- and *z*-direction. The resulting motion has been characterized by videos (see Supporting Information S2) recorded through the microscope with a high-speed camera (800 x 600 pixel resolution and 1000 fps frame rate). Again, the lithographically patterned resist structure on top of the EB substrates has been retained in order to relate particle position and orientation to the underlying domain pattern. In these experiments, an 8 nm thick NiFe layer was used for the substrate layer stack, which results in magnetic stray fields weak enough to not cause sticking of the particles to the substrate. (Note, that instead of this a 10 nm thick CoFe layer was used for the later discussed investigation on the influence of external field parameters. For these latter experiments, an additional PMMA layer was necessary to weaken the stray fields.) A trapezoidal, temporally periodic magnetic field sequence (see brown, dot-dashed and black, dotted lines in Figure 3(c)) was produced by using orthogonally placed Helmholtz coil pairs, where an alteration between $H_{z,max}$ ($H_{x,max}$) and $-H_{z,max}$ ($-H_{x,max}$) was induced every half period at an alteration rate of $3.2 \cdot 10^6$ Am$^{-1}$s$^{-1}$. As in previous experiments[32,34,36], a phase shift of $\pi/2$ between $H_z$ and $H_x$ has been used for the directed movement of the MJPs in +*x*- or –*x*-direction. When applying the magnetic field sequence, the exchange-biased MJPs dispersed in distilled water on top of an EB substrate with (hh)/(tt) stripe domains perform a combination of distinct translational and rotational movements (see Figure 3(a) (A-G)).

At $t = 0$ s, $-H_{z,max}$ and $-H_{x,max}$ are applied, positioning the particle close to a (tt) domain wall (Figure 3 (a) A). $H_x$ induces a slight shift of the potential energy minimum's location from the domain wall center towards the +*x*-direction, easily seen in Figure 3 (a) A.

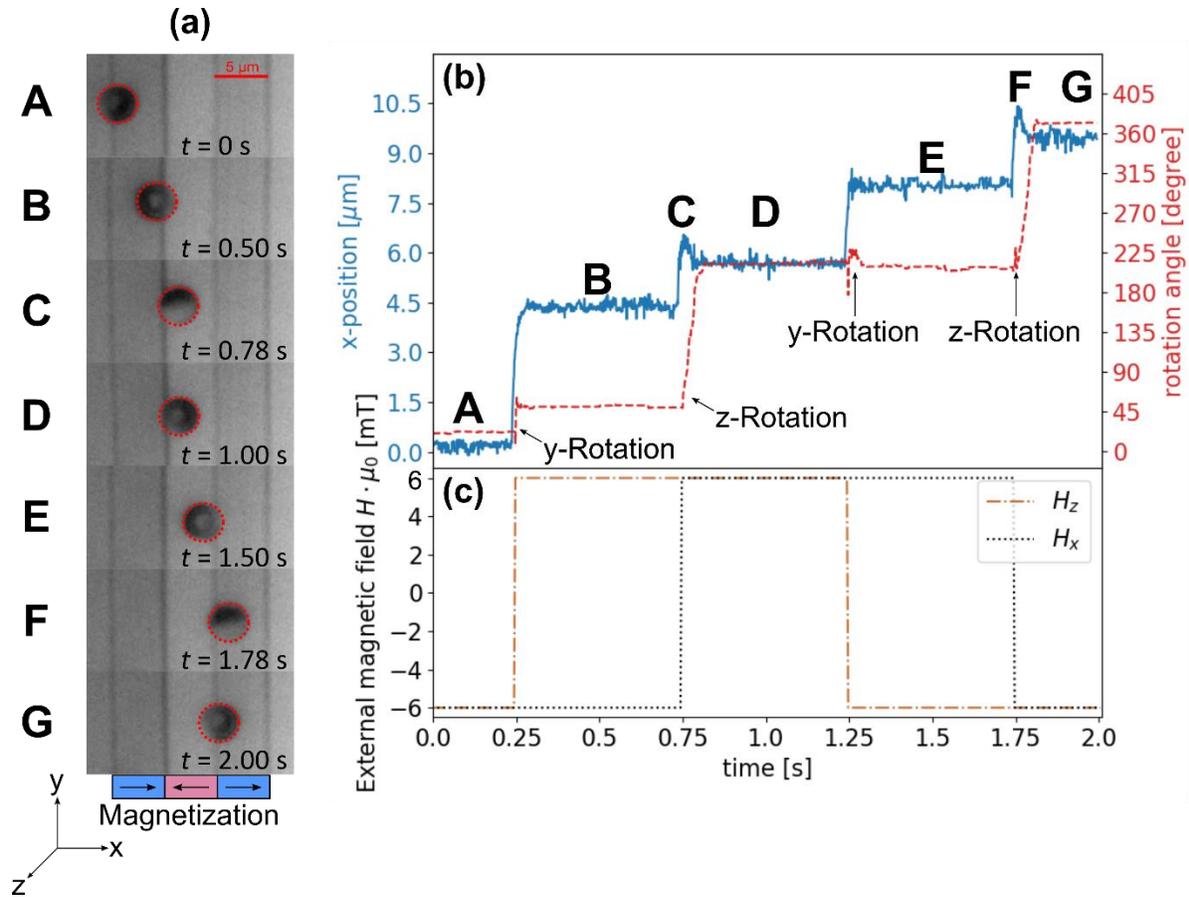

**Figure 3**. Directed transport of exchange-biased, magnetic Janus particles (MJPs) by the superposition of parallel-stripe static magnetic stray field landscapes and weak external magnetic field pulses. (a) Microscopy snapshots (A-G) for the different phases of the externally applied magnetic field pulse sequence (c). (b) Position (blue solid line) and cap orientation with respect to the z-axis (red, dashed line) determined from particle tracking in the respective videos. The stripe structure, optically visible in the microscopic images is the lithographically patterned resist left behind to mark the positions of the magnetic domains within the underlying substrate. The tracking data reflects the visually observable fast translational and rotational motion of the MJP induced by a change of the external magnetic field in *z*- (brown, dot dashed line) and *x*-direction (black, dotted line).

Another consequence of the applied external magnetic fields $-H_{z,max}$ and $-H_{x,max}$ is the spatial alignment of the MJP's magnetic cap along the resulting effective magnetic field vector, orienting the cap as sketched in Figure 4(a). At $t = 0.25$ s, the external field in z-direction is inverted, yielding $+H_{z,max}$ and $-H_{x,max}$ as amplitudes for the present external magnetic fields. Consequently, the potential energy landscape is transformed (see Figure 4(b)) and the MJP jumps accordingly towards the position of the adjacent (hh) domain wall in +x-direction, as it can be seen in Figure 3 (a) B. The effective magnetic field vector has rotated around the y-axis by 90° with a corresponding rotation of the particle's cap aligning the particle's magnetic moment along the direction of the effective magnetic field (as depicted in Figure 4(b)).

The dynamics of the MJP motion has been analyzed by tracking the particle's position in x-direction and the rotation angle of the MJP around the z-axis projected onto the substrate plane[37] (blue, solid respectively red, dashed line in Figure 3(b)). Changing $H_z$ results in a fast translational and rotational movement of the MJP occurring within a few microseconds (between 10 ms to 20 ms). Note that after a change of $H_z$ the rotation angle of the MJP by recording the projection of the cap orientation onto the substrate plane is due to a rotation around the y-axis not always unambiguous, hampering the accurate tracking of the MJP's cap orientation. The situation becomes different at $t = 0.75$ s, where a sign change of the magnetic field in x-direction from $-H_{x,max}$ to $+H_{x,max}$ occurs. This causes a particle motion over a short distance in +x-direction and a cap reorientation. In accordance with the calculated potential energy of the particle for the given magnetic field configuration (Figure 4(c)), the particle is now located above an (hh) domain wall, with a shift into the +x-direction with respect to the wall center (Figure 3 (a) D). The particle has been found to rotate around the z-axis in this case (see Figure 3 (a), B-D). In the particular case of Figure 3 the rotation sense has been clockwise, however, in general the sense of rotation is energetically degenerate.

At $t = 1.25$ s, the magnetic field in $z$-direction is inverted back to $-H_{z,max}$, inducing particle movement to the location of the adjacent (tt) domain wall (Figure 3 (a) E), with a slight shift in $-x$-direction with respect to the domain wall center. Again, the change of the effective magnetic field vector provokes a reorientation of the particle's magnetic cap (see Figure 4 (d)) and, thus, the observed reaction is a fast rotational movement around the y-axis. At $t = 1.75$ s, the initial configuration of $-H_{z,max}$ and $-H_{x,max}$ is regained. Consequently, the MJP is located again above a (tt) domain wall with a slight shift towards the $+x$-direction (Figure 3 (a) G) with a spatial reorientation of the particle's cap by a rotation around the z-axis (highlighted in Figure 3 (a) F).

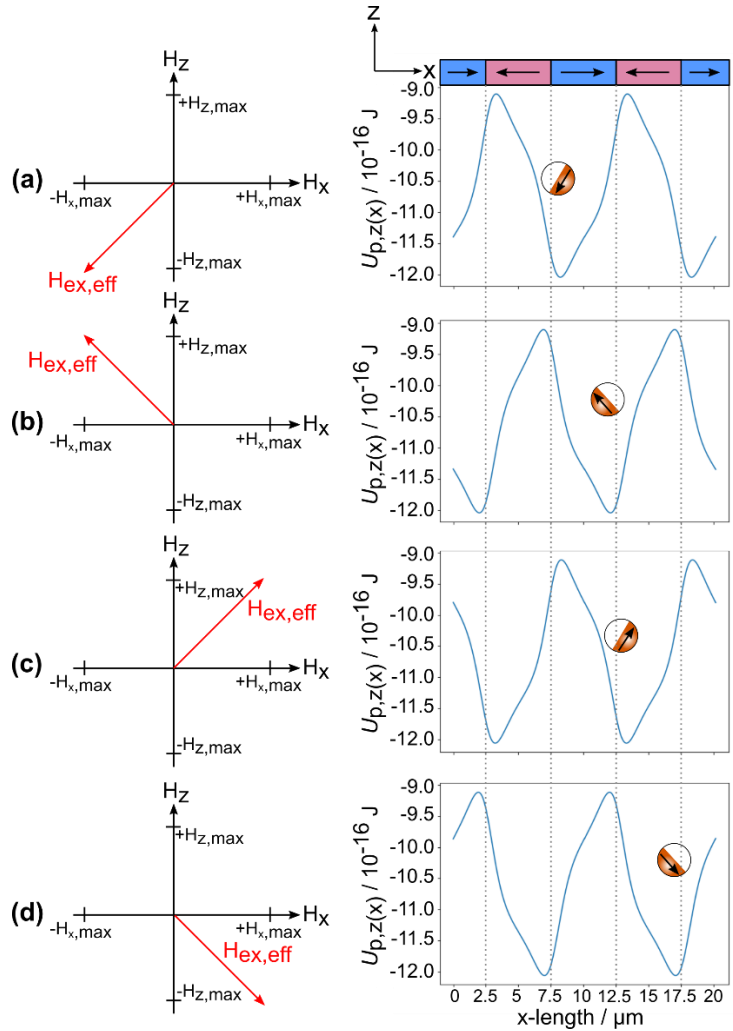

**Figure 4.** Correlation of effective external magnetic field vector (left panels), spatial position and orientation of exchange-biased, magnetic Janus particles (MJPs) (sketches in right panels) and the calculated potential energy landscape $U_{P,z}(x)$ (solid, blue lines in right panels). Each configuration for the applied external magnetic field sequence used in the present transport experiments is depicted (rows (a) – (d)).

The described transport of exchange biased MJPs is similar to the directed movement of superparamagnetic beads (SPB) over the same magnetic pattern and employing the same external magnetic field sequence[34]. For superparamagnetic beads, however, a reorientation of the particles' magnetic moment by an external magnetic field either occurs by Néel (pure spin reorientation) or Brownian relaxation (reorientation by physical rotation)[38]. For the present MJPs with a spatially fixed remnant magnetic moment in the exchange-biased cap (magnetic onion state[16]) only Brownian relaxation is possible as long as the external field does not induce a magnetization reversal in the particle. As the cap is optically opaque its direction has been determined from the recorded videos by an automated method employing machine learning[37]. Thus, a full characterization of the particle's translatory and rotatory motion is possible. By modifying the external field pulse sequence (frequency, directions, magnitudes) particle rotations may be induced such that all surface elements of the particle will face the substrate surface. This is of high importance when a small number of analytes on the surface of the particle shall be detected by formation of analyte bridges to the substrate and corresponding immobilization of the particles.

The MJPs possess a comparably high magnetic net moment of the metallic cap and therefore exhibit fast motion dynamics, as is understandable from the expression for the particle steady-state velocity $\vec{v}_P(x,z)$ [34]:

$$\vec{v}_P(x,z) = -\frac{\mu_0 \cdot (\vec{m}_P \cdot \vec{\nabla}) \cdot \vec{H}_{\text{eff}}(x,z)}{3 \cdot \pi \cdot d_P \cdot \eta_D \cdot f_D(z)}. \tag{4}$$

Here, Stokes law for low Reynolds number laminar flow describes the drag force acting on the particles, with $d_P$ being the hydrodynamic particle diameter, $\eta_D$ the viscosity of the surrounding fluid and $f_D$ the drag coefficient[26]. It becomes obvious, that high steady-state velocities can be achieved by high magnetic moments $\vec{m}_P$ of the used particles and a high gradient of the effective local magnetic field $\vec{H}_{\text{eff}}(x,z)$. Both conditions are fulfilled here. This is corroborated by the experimentally determined MJP transport velocity (Figure 3). Steady-state transport velocities were retrieved by fitting a Gaussian error function to the obtained MJP trajectories after each alteration of $H_z$ within a time frame of $T/4$ (with $T$ being the period of the external magnetic field pulse sequence) and extracting the time derivative. This results in a Gaussian distribution of velocities, with the instantaneous particle step velocity taken as the maximum of the curve. Averaging over all observed transport steps after a change of the external $H_z$, the instantaneous steady-state transport velocity amounts to $(214 \pm 8)$ µm/s. Compared to previously reported translatory motion speeds of remotely controlled MJPs[18,19,22,24], this velocity is at least one order of magnitude higher.

Additionally we have investigated the MJP transport and rotation properties as functions of the pulse plateau time length $\Delta t$ (Fig. 5 (a)) and the strengths of the applied field components in z- and x-direction $\Delta H$ (Fig. 5 (c)). For pulse plateau times shorter than a critical time $\Delta t_{c,t}$ for the translatory motion no transport is observed. Here, the MJPs cannot follow the fast changes of the potential energy landscape[34]. For plateau times longer than $\Delta t_{c,t}$ (in the current experiment $\Delta t_{c,t} \approx 100$ ms) the instantaneous transport velocity for different pulse plateau time lengths (blue triangles in Figure 5(b)) is constant within the experimental uncertainties, (see Eq. 4), similar to previous results for transport of superparamagnetic beads[34]. The rotation properties have been characterized by determining the angle over which the cap re-orientates (grey circles in Figure 5(b)), for a rotation around the z-axis after a sign change of the external $H_x$-field (see Figure 3(b)).

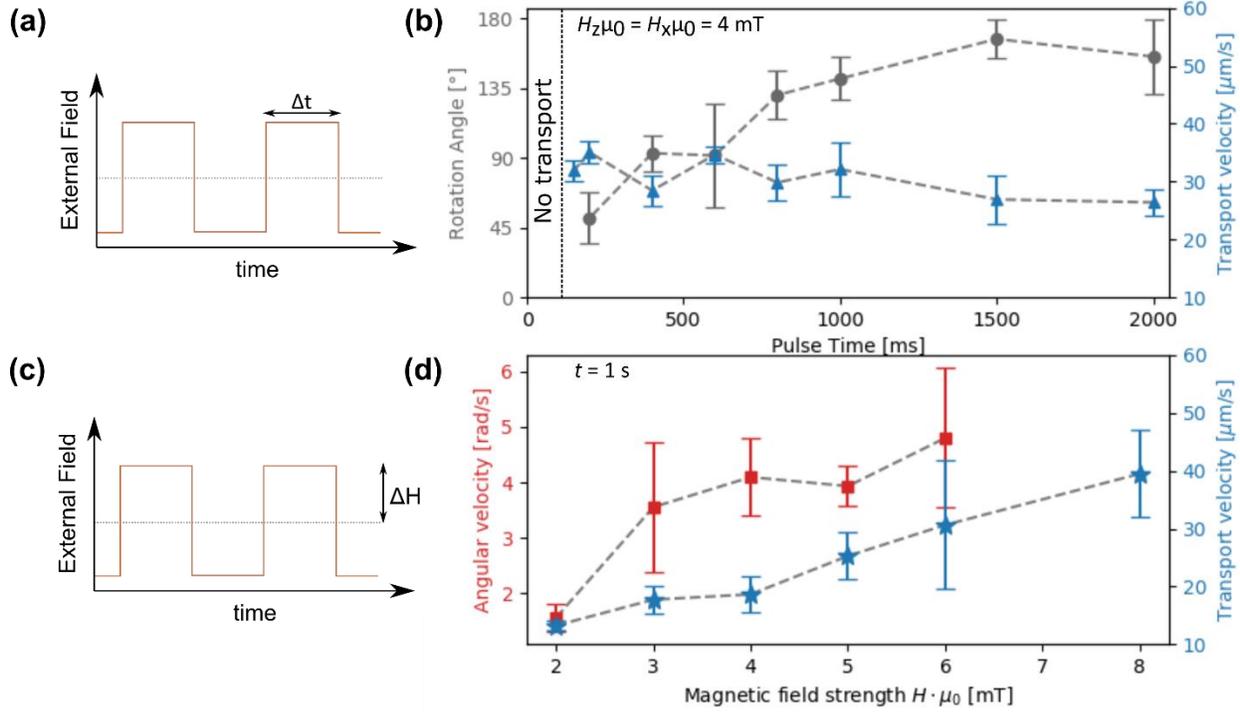

**Figure 5**. Quantitative analysis of the motion characteristics of exchange-biased, magnetic Janus particles (MJPs) in dependence of the pulse plateau time length $\Delta t$ (a) and the field strength $\Delta H$ (c) of the externally applied magnetic field. Experiments have been performed using a continuous 2400 nm thick PMMA capping layer on the magnetic film system to avoid particle sticking. In panel (b) blue triangles indicate the instantaneous steady-state transport velocity determined as described in the text, while grey circles symbolize the measured angle determined for the accompanying rotation around an axis perpendicular to the substrate plane (z-axis). Panel (d) shows the instantaneous steady-state transport velocity in blue stars and the instantaneous steady-state angular velocity (for a rotation around the z-axis) in red squares. Transport velocities were determined in dependence of $\Delta H_z$ and angular velocities were determined in dependence of $\Delta H_x$ (while keeping the respective other $\Delta H$ constant). Dashed lines serve as guides to the eye,

error bars mark the standard deviation determined from the number of investigated motion events for one observed particle.

For $\Delta t \geq \Delta t_{c,r} \approx 1500$ ms a complete cap reorientation by 180° has been observed during the pulse plateau with a statistical distribution of clockwise or counterclockwise rotations. For $\Delta t < \Delta t_{c,r}$ the cap rotation angle decreases, indicating that there is not enough time for a complete cap reorientation prior to the subsequent inversion of $H_z$. At $\Delta t \leq 150$ ms a cap rotation was not observable at all. We may, therefore, define a critical pulse plateau time $\Delta t_{c,r}$ above which a cap reorientation by 180° occurs, similar to $\Delta t_{c,t}$. The difference between these two critical pulse plateau time lengths is important for an individual control of translatory and rotatory motion in lab-on-a-chip devices. When discussing the time-averaged translation and rotation velocities in contrast to the in this work determined instantaneous steady-state velocities, we may distinguish the following cases:

a) $\Delta t_{c,r} < \Delta t_{c,t}$:

   1) $\Delta t \ll \Delta t_{c,r} < \Delta t_{c,t}$: MJPs are translating/rotating with the frequency of the varying external field in the non-linear regime[39,40], with a decreasing time-average of the angular and linear velocities with decreasing $\Delta t$ for rotation and translation, respectively. For very small $\Delta t$ only a lateral forward/backward and rotational clockwise/counterclockwise, oscillation will occur.

   2) $\Delta t_{c,r} < \Delta t < \Delta t_{c,t}$: Full cap reorientations will occur, superposed on the lateral non-linear transport regime[39].

3) $\Delta t_{c,r} < \Delta t_{c,t} < \Delta t$: Both, lateral transport and rotation are in the linear regime[39,40], where a full step is observed for each field pulse accompanied with full cap reorientations in each transport step (as shown in Figure 3).

b) $\Delta t_{c,t} < \Delta t_{c,r}$:

1) $\Delta t < \Delta t_{c,t} < \Delta t_{c,r}$: Same as (a1).

2) $\Delta t_{c,t} < \Delta t < \Delta t_{c,r}$: Directed lateral transport in the linear regime[39] accompanied by rotation in the non-linear regime[40].

3) $\Delta t_{c,t} < \Delta t_{c,r} < \Delta t$: Same as (a3).

We now discuss some possibilities to experimentally modify the two critical time intervals: for $\Delta t_{c,t}$ one trivial experimental parameter is the ratio between the domain widths of the underlying parallel-stripe domains and the pulse plateau time length. This has been studied in this case for 5µm wide parallel stripes and varying the pulse plateau time length at given field strengths of the applied magnetic field components (Figures 5 (a), (b)). Here, a critical pulse plateau time length of 100 ms for transport has been observed. The critical time interval for a full cap reorientation $\Delta t_{c,r}$ is of course independent of the parallel-stripe domain widths and has been determined here to be about 1500 ms (Figure 5(b)). In this case however, the applied magnetic field strengths $H_z$ and $H_x$ (depending on the rotation axis) may be influential. Further studies on modifying $\Delta t_{c,t}$ and $\Delta t_{c,r}$ will be carried out in future works in order to determine additional crucial experimental parameters. Finally, we have investigated the influence of different magnitudes of the applied field on the angular and lateral transport velocities of the MJPs. As can be seen in Figure 5 (d), the lateral velocity increases almost linearly with increasing field component $\Delta H_z$ (keeping $\Delta H_x$ constant) and the angular velocity also increases with increasing $\Delta H_x$ (keeping $\Delta H_z$ constant). Both observations can be explained by the increasing magnetic force and torque on the MJP.

*Conclusion:*

Making use of tailored magnetic stray fields above a prototypical, artificial magnetic domain pattern superposed by a defined magnetic field pulse sequence, we demonstrated a controlled combination of a fast translatory and rotatory motion for spherical magnetic Janus particles (MJPs) on top of a topographically flat substrate in aqueous solution. The MJPs were engineered for having a comparably large and spatially fixed remanent magnetic moment by depositing an in-plane magnetized, exchange-biased thin film cap on one half of non-magnetic silica spheres. By applying weak external magnetic field pulses in the range of a few mT a stepwise motion pattern of the particles has been achieved. The motion pattern consists of a stepwise translation in one direction accompanied by MJP rotations around two distinct axes depending on the configuration of the effective external magnetic field vector. The addressability of the MJP's spatial orientation due to realignment of the half shell's magnetic moment with the effective magnetic field vector while being transported, makes this kind of motion behavior potentially applicable for enhancing biomolecule interactions in future lab-on-a-chip devices. Owing to high magnetic field gradients within locally defined stray fields and large magnetic moments of the used MJPs, instantaneous transport velocities of up to 200 µm/s can be achieved. This is at least one order of magnitude higher than previously reported motion speeds for other types of MJPs. Varying the chosen pulse plateau time length and amplitude of the applied external magnetic field sequence, different motion regimes of the particles can potentially be addressed: either a distinct combination of particle transport and rotation or an individual appearance of one of these motion types. Consequently, a change of just one external parameter is sufficient to induce the desired motion behavior of the MJP, thus adding additional flexibility towards specific applications and separately controlled degrees of freedom.

*Methods:*

Exchange biased Janus particles (JPs) were fabricated by sputter depositing a $Cu^{5\,nm}/Ir_{17}Mn_{83}^{30\,nm}/Co_{70}Fe_{30}^{10\,nm}/Si^{10\,nm}$ layer system on top of spherical silica beads with a diameter of $d = 3$ μm (sicastar, micromod Partikeltechnologie GmbH). Prior to that, the silica beads were assembled in a monolayer on a glass substrate (1 cm x 1 cm), which was cleaned for 24 h in concentrated sulfuric acid, according to the method introduced by Micheletto et al[13]. After evaporation of the dispersion fluid the glass substrate was transferred to a rf sputter machine for depositing the magnetic cap onto the particles. During the deposition (working pressure = $10^{-2}$ mbar) a homogenous in-plane magnetic field of 35 mT was applied in order to set the exchange bias (EB).

Transport experiments were conducted in a home-built setup consisting of a combination of Helmholtz coils for the application of magnetic field pulses in *x*- and *z*-direction, an optical bright field microscope with a 100x magnification objective (Nikon, N.A. = 1.4) and a high speed camera (Optronis CR450x2) for recording videos of the particle dynamics with 1000 frames per second and a resolution of 800 x 600 pixels. In preparation for the transport experiments the JPs were collected from the glass substrate either by mechanical scratching and dispersing in distilled water or by sonication of the substrate in distilled water. Before applying the particles to the experimental setup, the acquired dispersions were sonicated for ca. 10 min to reduce agglomeration of particles. A volume of 30 µL of the JP dispersion was pipetted into a microfluidic chamber, which was created by cutting a window of approximately 10 x 10 mm into a Parafilm® sheet attached to the magnetically stripe patterned ((hh)/(tt) domain configuration) EB substrate with a size of ca. 15 x 15 mm. After adding the particle dispersion to the chamber (side wall height = ca. 100 µm) it was sealed with a coverslip (Carl Roth), on which a drop of immersion oil (AppliChem GmbH, A0699,0100) was placed. The whole arrangement was put onto a substrate holder in the middle of the Helmholtz coil setup with the substrate plane being parallel to the direction of the magnetic field generated by the *x*-coils, i.e. the magnetic field in *z*-direction is perpendicular to the substrate plane. For inducing particle motion, trapezoidal magnetic field pulses were applied with a temporal phase shift of $\pi/2$ between *x*- and *z*-pulses. Each half period of these pulses consisted of a linear rising time for the magnetic field, a plateau time and a linear drop time. The rising and drop times are given by the alteration rate of the magnetic field ($3.2 \cdot 10^6$ $Am^{-1}s^{-1}$) and the applied field strength.

The transport substrates were fabricated by applying ion bombardment induced magnetic patterning (IBMP) to a field cooled $Cu^{5\ nm}/Ir_{17}Mn_{83}^{30\ nm}/Co_{70}Fe_{30}^{10\ nm}$ ($Ni_{80}Fe_{20}^{8\ nm}$)$/Au^{10\ nm}$ EB layer system. After depositing the layer system via rf sputtering onto a naturally oxidized Si(100)-Wafer a subsequent field cooling procedure was applied in order to set the EB direction. Here, the substrates were annealed in a vacuum chamber (base pressure = 5 x $10^{-7}$ mbar) at 300 °C for 60 min while an in-plane magnetic field of 145 mT was applied. For magnetically patterning the EB layer system a photoresist was spin coated on top of the substrates and 5 µm wide stripe structures (periodicity of 10 µm) perpendicular to the field-cooled induced EB direction were created by photolithography (Karl Suss MA-4 Mask Aligner). Subsequent to the application of this shadow mask, the samples were bombarded with He ions (10 keV) in a home-built setup consisting of a custom Penning ion source. During bombardment an in-plane magnetic field of 80 mT was present, with the direction of the magnetic field being anti-parallel to the field-cooled induced EB direction to achieve a hh-tt domain configuration. For the conducted quantitative studies, a topographically flat PMMA layer was fabricated on top of the substrate in order to weaken the magnetic force acting on the JPs and thereby to reduce extensive sticking of the particles to the substrate. Consequently, lower transport velocities than presented for the qualitative investigation (Figure 3) are measured, making it however easier to probe the influence of the mentioned experimental parameters. Therefore, the photoresist was removed by treating the samples gradually in an ultrasonic bath for 5 min at 50 °C in a 3% KOH solution and for 3 min at 50 °C in acetone and water, respectively. At last, the samples were cleaned with acetone, isopropanol and water and dried in a $N_2$ stream. A 2.4 µm thick PMMA layer was spin-coated on top of EB substrates before conducting transport experiments.

ACKNOWLEDGMENT

The authors thank the Center for Interdisciplinary Nanostructure Science and Technology (CINSaT) at Kassel university promoting cross-disciplinary communication and research, as well as project "MASH" supported by an internal grant of Kassel university.

ABBREVIATIONS

LOC = lab-on-a-chip

µTAS = micro-total-analysis-system

MJP = magnetic Janus particle

EB = Exchange-Bias

IBMP = ion bombardment induced magnetic patterning

SPB = superparamagnetic beads

6. Campuzano, S. *et al.* Magnetic Janus Particles for Static and Dynamic (Bio)Sensing. *Magnetochemistry* **5**, 47 (2019).

7. Yi, Y., Sanchez, L., Gao, Y. & Yu, Y. Janus particles for biological imaging and sensing. *Analyst* **141**, 3526–3539 (2016).

8. Lee, K., Yi, Y. & Yu, Y. Remote Control of T Cell Activation Using Magnetic Janus Particles. *Angew. Chemie Int. Ed.* **55**, 7384–7387 (2016).

9. Moerland, C. P., Van IJzendoorn, L. J. & Prins, M. W. J. Rotating magnetic particles for lab-on-chip applications-a comprehensive review. *Lab on a Chip* **19**, 919–933 (2019).

10. Walther, A. & Müller, A. H. E. Janus Particles: Synthesis, Self-Assembly, Physical Properties, and Applications. *Chem. Rev.* **113**, 5194–5261 (2013).

11. Güell, O., Sagués, F. & Tierno, P. Magnetically driven Janus micro-ellipsoids realized via asymmetric gathering of the magnetic charge. *Adv. Mater.* **23**, 3674–3679 (2011).

12. Hong, L., Jiang, S. & Granick, S. Simple method to produce janus colloidal particles in large quantity. *Langmuir* **22**, 9495–9499 (2006).

13. Micheletto, R., Fukuda, H. & Ohtsu, M. A Simple Method for the Production of a Two-Dimensional, Ordered Array of Small Latex Particles. *Langmuir* **11**, 3333–3336 (1995).

14. Streubel, R. *et al.* Equilibrium magnetic states in individual hemispherical permalloy caps. *Appl. Phys. Lett.* **101**, 132419 (2012).

15. Albrecht, M. *et al.* Magnetic multilayers on nanospheres. *Nat. Mater.* **4**, 203–206 (2005).

**Supporting Information:**

**S1: Calculation of steady-state distances**

For the calculation of steady-state distances between Janus particles and the surface of the underlying substrate, three relevant forces are balanced against each other, depending on whether they are attractive or repulsive forces. We take the following forces into account: the magnetic force $\vec{F}_M(z)$ (see Eq. (2) in the manuscript) between the magnetic substrate and the magnetic cap of the particle, the van der Waals force $\vec{F}_{vdW}(z)$ originating from interactions between dipoles and/or induced dipoles and the electrostatic force $\vec{F}_{el}(z)$. Employing the Hamaker constant $A_{132}$ for a particle of material 1 above a substrate of material 2 surrounded by a medium of material 3, the van der Waals force can be expressed as follows:

$$\vec{F}_{vdW}(z) = -\frac{A_{132} \cdot R_P}{6z^2} \cdot \left[\frac{1}{1 + 14\frac{z}{\lambda_{ret}}}\right] \cdot e_z, \qquad (5)$$

With $\lambda_{ret}$ representing the retardation wavelength of the interaction, which in this case is assumed to be 100 nm. Considering the immersion of the particles and the substrate surface in an ionic liquid like water, a double-layer of opposite charges is forming at each surface, leading to an electrostatic force between the surfaces when their respective double-layers start to overlap. This force is governed by the surface potentials of the particle $\Psi_P$ and the substrate $\Psi_S$, respectively, and the inverse double-layer thickness κ, given by the Debye-Hückel theory. Now, the electrostatic force can be written as:

$$\vec{F}_{el}(z) = \frac{2 \cdot \pi \cdot \varepsilon \cdot \kappa \cdot R_P}{1 - e^{-2\kappa z}} \cdot [2 \cdot \Psi_S \cdot \Psi_P \cdot e^{-\kappa z} \mp (\Psi_S^2 + \Psi_P^2) \cdot e^{-2\kappa z}], \qquad (6)$$

with ε being the permittivity of the medium. Depending on the used sign in Eq. (2), it can be differentiated between a model for the electrostatic force, where the surface potentials of both particle and substrate are assumed to be constant (upper sign) and where the surface charge density of both components is assumed to be constant (lower sign). Since both of these models describe extreme cases for the behavior of the here studied system, calculations for the electrostatic force were conducted using both approaches. In order to obtain the steady-state distance between substrate surface and Janus particle, $\vec{F}_M(z), \vec{F}_{vdW}(z)$ and $\vec{F}_{el}(z)$ were computed in dependence of $z$ for particles with a diameter of 3 µm. For the calculation of $\vec{F}_M(z)$, the simulated magnetic stray field landscape $\vec{H}_{MFL}(x,z)$ above the substrate (retrieved from micromagnetic simulations for the used domain configuration) was considered at the position $x$ above the center of a domain wall. For $\vec{F}_{vdW}(z)$, a Hamaker constant of $A_{132} = 3.4 \cdot 10^{-21}$ was chosen, leading to an attractive van der Waals interaction between particle and substrate. However, in the case of $\vec{F}_{el}(z)$ a repulsive force is present, since for $\Psi_S$ and $\Psi_P$ the zeta potentials of PMMA (-35 mV) and glass (-35 mV) at pH = 7 were used, respectively. Balancing the sum of $\vec{F}_M(z)$ and $\vec{F}_{vdW}(z)$ in dependence on the distance between particle surface and resist surface and the repulsive force $\vec{F}_{el}(z)$ yields the stead-state distance. It is important to note, that the gravitational force $F_G$ and the buoyancy force $F_B$ were not considered for the calculation of the steady-state distance, since they are two to three orders of magnitude smaller for Janus particles with $d = 3$ µm than the discussed forces which are in the range of $10^{-10}$ N and $10^{-12}$ N for the relevant ranges of $z$. When finding that both models for the electrostatic force yield two clear intersections with the attractive forces curve, we take the average of both as the estimated steady-state distance.